\documentclass[12pt]{article}
\usepackage[colorlinks=true, pdfstartview=FitV, linkcolor=blue, citecolor=blue, urlcolor=magenta]{hyperref}
\usepackage{eurosym}
\usepackage{dcolumn}
\usepackage{bm}

\newcommand{\beq}{\begin{equation}}
\newcommand{\eeq}{\end{equation}}
\newcommand{\bea}{\begin{array}}
\newcommand{\eea}{\end{array}}
\newcommand{\bey}{\begin{eqnarray}}
\newcommand{\pslash}{\not{\hbox{\kern-2.3pt $p$}}}
\newcommand{\pdslash}{\not{\hbox{\kern-2pt $\partial$}}}

\newcommand{\eey}{\end{eqnarray}}
\begin{document}

\begin{titlepage}

\vskip 2cm
\begin{center}
{\Large\bf   Lorentz-violating extension of the spin-one Duffin-Kemmer-Petiau equation \footnote{{\tt belich@cce.ufes.br}, {\tt passos@df.ufcg.edu.br}, {\tt mdemonti@ualberta.ca}, {\tt esdras.santos@ufba.br} }}
 \vskip 10pt
{\bf
  H. Belich$^{a}$, E. Passos$^b$, M. de Montigny$^c$, E. S. Santos$^{c,d}$  \\}
\vskip 5pt
{\sl $^a$ Departamento de F\'{\i}sica e Qu\'{\i}mica\\
 Universidade Federal do Espirito Santo, Av. Fernando Ferrari, \\
S/N, Goiabeiras, Vit\'oria, ES, 29060-900, Brazil}\\
{\sl $^b$ Departamento de F\'{\i}sica, Universidade Federal de Campina Grande\\  
58109-970 - Campina Grande, PB, Brazil}\\
{\sl $^c$Facult\'e Saint-Jean, University of Alberta, \\
Edmonton, Alberta, Canada T6C 4G9\\}
{\sl $^d$Instituto de F\'{\i}sica, Universidade Federal da Bahia\\
40210-340 - Salvador, BA, Brazil\\}
\vskip 2pt
\end{center}

\begin{abstract}
We investigate the breaking of Lorentz symmetry caused by the inclusion of an external four-vector via a Chern-Simons-like term in the Duffin-Kemmer-Petiau Lagrangian for massless and massive spin-one fields. The resulting equations of motion lead to the appearance of birefringence, where the corresponding photons are split into two propagation modes. We discuss the gauge invariance of the extended Lagrangian. Throughout the paper, we utilize projection operators to reduce the wave-functions to their physical components, and we provide many new properties of these projection operators. 
\end{abstract}

\bigskip


{\em PACS:} 3.65.Ge; 11.10.Nx; 11.10.Kk; 11.10.-z

\end{titlepage}  

%%%%%%%%%%%%%%%%%%%%%%%%%%%%%%%%%%%%%%%%%%%%%%%%%%%%%%%%%%%%%%%%%%%%%%%

%%%%%%%  1 -  Introduction

\section{Introduction\label{introduction}}

The brilliant success of the Standard Model (SM) of elementary particles  is
still hampered by some hurdles. For instance, the SM
has not been successful in explaining the origin of electron's electric
dipole moment, $d_{e}$, and its experimental upper bounds \cite{revmod,science}.
 This motivates investigations of physics
beyond the SM. Along these lines, a possible way is to extend the mechanism of the
spontaneous symmetry breaking through a background vector (or tensor) field
such that the Lorentz symmetry is violated \cite{ens}. In 1989, Kosteleck\'{y} and
Samuel \cite{extra3} discussed an interesting possibility of establishing
the spontaneous violation of symmetry through non-scalar field (vacuum of
fields that have a tensor nature) based on a string field theory
environment. A general framework for testing the low-energy manifestations
of CPT violation and Lorentz symmetry breaking is the Standard-Model
Extension (SME) \cite{colladay-kost}, where the effective Lagrangian
corresponds to the usual SM Lagrangian, to which are added SM operators of
any dimensionality contracted with Lorentz-violating tensorial background
coefficients \cite{coll-kost,baeta,bras}. With regard to the experimental searches for
CPT- and Lorentz-violation, the generality of the SME has provided the basis
for many investigations. In the flat spacetime limit, empirical studies
include muons \cite{muon}, mesons \cite{meson,meson2}, baryons \cite{barion,barion2}, photons \cite{photon,photon1,petrov}, electrons \cite{electron},
neutrinos \cite{neutrino} and the Higgs sector \cite{higgs}. The gravity
sector has also been explored in Refs. \cite{gravity,gravity2}.  Current limits on the coefficients of the Lorentz
symmetry violation are in Ref. \cite{data}. In recent years, Lorentz symmetry breaking effects have
been investigated in the hydrogen atom \cite{manoel}, in the Rashba coupling 
\cite{rash,bb3}, in a quantum ring \cite{bb4}, in Weyl semi-metals \cite{weyl}, in tensor backgrounds \cite{louzada,manoel2}, in the quantum Hall
effect \cite{lin2} and geometric quantum phases \cite{belich,belich1,bbs2}.

In this paper, we will apply a similar Lorentz-violation approach to the
Duffin-Kemmer-Petiau (DKP) equation \cite{dkp,Umezawa,Corson}. Originally intended to describe
mesons, and sometimes called the `meson algebra', the DKP theory describes
massive \cite{dkp} and massless \cite{harish-chandra} scalar and vector
bosons in a unified formalism based on a first-order wave equation
analogous to the Dirac equation for spin-half fields. Hence, for spin-zero
bosons, one replaces the second-order Klein-Gordon equation with the
first-order DKP equation which involves matrices $\beta^\mu$, analogous to
the Dirac gamma matrices, that satisfy a specific algebraic relation such
that the DKP equation acquires a matrix form. 
Despite its similarity with the Dirac equation, the DKP formalism is more intricate; for instance, its field components are dependent, the use of specific representations can sometimes be replaced by component-projection operators, the treatment of massless fields requires more than simply setting the mass to zero and involves singular operators, etc. The DKP theory has been applied in different problems in quantum mechanics and field theory, such as the meson-nucleus elastic scattering \cite{clark}, quantum chromodynamics \cite{gribov}, covariant Hamiltonian dynamics \cite{kanatchikov}, studies on the $S$-matrix \cite{pimentel}, calculations of
the phase in Aharonov-Casher effect \cite{swansson}, on the causality of the
DKP theory \cite{lunardi}, Bose-Einstein condensation \cite{bec1, bec2},
curved spacetime \cite{casana}, non-relativistic theories via Galilei
covariant $5$-dimensional formalism \cite{galileana1, galileana2, galileana3}
and several other applications \cite{potenciais}.

With a view to implementing Lorentz-symmetry breaking, the DKP formalism is
interesting because it allows us to introduce various types of interactions
through scalar, pseudo-scalar, vector or tensor couplings \cite{guertin}.
The richness of couplings introduced in the DKP theory allows us to
examine Lorentz-symmetry breaking via non-minimal couplings of the
massless DKP field with a background field. 
In this paper, we add a Chern-Simons-type term in the DKP Lagrangian for
massless and massive spin-one fields, that causes the breaking of Lorentz
symmetry. As studied in the literature, a sensitive phenomenological
evidence of Lorentz-symmetry violation would be provided by the observation
of vacuum birefringence. Carroll, Field and Jackiw observed that, in $3+1$ dimensions, the Chern-Simons term, $n_\mu
A_\nu{\tilde F}^{\mu\nu}$, which couples the dual electromagnetic tensor to
an external (or background) four-vector (denoted $n_\mu$ hereafter) is gauge invariant but not
Lorentz invariant \cite{photon}.  One observes the birefringence of light in the vacuum
when speeds which depend on polarization appear in the solutions of the
modified Maxwell equations with Lorentz-violating terms (see also Kostelecky
and Mewes (2009, 2013) in Ref. \cite{photon1}). In this paper, we will point
out that our model implies a connection between the photon dispersion
relation and its polarization, which can therefore lead to a vacuum
birefringence effect.

We examine a modified theory of electromagnetism with focus
on the gauge sector of the SME. With a study based on the DKP formalism, we
carry out the analysis of the odd sector \cite{photon}. 
In Sec. \ref{masslessdkp}, we set up the model for spin-one massless DKP
fields and we modify it by adding a Lorentz-violating background
vector. In Sec. \ref{birefringence}, we obtain and analyse the dispersion
relations of the model. In Sec. \ref{massivedkp}, we study the massive DKP
equation with a Lorentz-violating term and obtain the dispersion
relation. Finally, we present concluding remarks in \ref{conclusion}.

%%  2 - DKP equation for massless field

\section{DKP equation for massless field in a background\label{masslessdkp}}

As mentioned in the introduction, one aspect of the DKP theory which is less
 straightforward than the Dirac equation is the treatment of massless fields, described in this section.
The Lagrangian for the massless DKP free field can be obtained in a manner similar to Harish-Chandra in
Ref. \cite{harish-chandra}. We write it as follows: 
\begin{equation}
{\mathcal{L}} = \frac{i}{2} {\overline{\Psi}} \beta ^{\mu} \partial _{\mu}
\Psi - \frac{i}{2} \left( \partial _{\mu} {\overline{\Psi}} \right) \beta
^{\mu} \Psi -{\overline{\Psi}}\gamma \Psi,  \label{L}
\end{equation}
where $\mu=0,1,2,3$, and ${\overline{\Psi}} = \Psi ^{\dag} \eta$ is the
adjoint DKP spinor, with $\eta = 2 (\beta ^0)^2-1$. 
We utilize the Minkowski metric $g^{\mu\nu}$ with signature $(+1, -1, -1, -1)$.

The corresponding free
massless DKP equation obtained from this Lagrangian is 
\begin{equation}
\left( i\beta^\mu \partial_\mu - \gamma \right) \Psi=0.  \label{dkp}
\end{equation}
Note the appearance here of
a singular matrix $\gamma$ which takes the place of the mass term,
 in contrast with the Dirac equation, where one simply takes the mass equal to zero. For vector DKP field with spin one, we take $\gamma =3-\beta^\mu\beta_\mu$. The
matrices $\beta^\mu$ and $\gamma$ satisfy the following algebra 
\begin{eqnarray}
\beta^\mu \beta^\nu \beta^\rho+\beta^\rho \beta^\nu \beta^\mu&=&
g^{\mu\nu}\beta^\rho+g^{\nu\rho}\beta^\mu,  \label{algebra1} \\
\gamma\beta^\mu+\beta^\mu\gamma&=&\beta^\mu,\;\;\;\;\;\gamma^2=\gamma.
\label{algebra2}
\end{eqnarray}

The DKP field, $\Psi$, has two different sectors: the scalar (or spin-zero)
sector whose representation is by $5\times 5$ beta-matrices, and the
vector spin-one sector represented by $10\times 10$ beta-matrices. 
The scalar DKP equation is equivalent to the Klein-Gordon equation,
 whereas the vector DKP equation with mass corresponds to the Proca field and to
  the Maxwell equation for massless fields. In both cases, the DKP equation
   provides rich possibilities to include interactions. Note that hereafter,
    we will not consider external sources, that is, we consider $j^\mu=0$.

The Lorentz-symmetry breaking is produced by the addition of a background
field to the Lagrangian, Eq. (\ref{L}), analogous to the Chern-Simons term
utilized in the literature. We write the DKP Lagrangian with these
symmetry-breaking terms in the form
\begin{eqnarray}
{\mathcal{L}} &=& \frac{i}{2} {\overline{\Psi}} \beta ^{\mu} \partial _{\mu}
\Psi - \frac{i}{2} \left( \partial _{\mu} {\overline{\Psi}} \right) \beta
^{\mu} \Psi -{\overline{\Psi}}\gamma \Psi -\frac{1}{4}{\overline{\Psi}}
\epsilon_{\mu\nu\rho\sigma}P\left[\beta^\mu,\;\beta^\nu\right]n^\rho\partial^\sigma\Psi  \nonumber \\
& &+\frac{1}{4}{\overline{\Psi}} \epsilon_{\mu\nu\rho\sigma}P\left[\beta^\mu,\;\beta^\nu\right]n^\rho\overleftarrow{\partial}^\sigma\Psi,  \label{L1}
\end{eqnarray}
where, for the vector DKP field with spin one, we have 
\begin{equation}
P=1-\gamma.\label{P}
\end{equation} 
The model is CPT-odd and predicts a rotation of the plane of polarization of radiation from distant galaxies such as in gamma-ray emission \cite{photon}. The other contribution to the pure-photon sector is a CPT-even Lorentz-violating term which does not have such properties \cite{coll-kost}.
For the scalar sector, the projection operator $P_S$ is defined in Appendix \ref{AppA}.
The four-vector $n^\mu$ is constant and acts as the background which breaks the
Lorentz symmetry (see Ref. \cite{photon}). 
(Note that the commutator $\left[\beta^\mu,\;\beta^\nu\right]$, which is in the definition of the spin operator {\bf S},
\begin{equation}
S_j=\frac i2\epsilon_{jkl}\left[\beta^k,\beta^l\right],\label{spinoperator}
\end{equation}
therefore appears in the expression for the Pauli-Lubanski vector, which gives the field's spin \cite{luciano}. This suggests that the new interaction term in the Lagrangian should be trivial for spin-zero fields, as we will explain below Eq. (\ref{dkp-adjunta}).)

The interaction term between the DKP field $\Psi$ and the background field $n^\rho$ in Eq. (\ref{L1}) is related to the Chern-Simons term employed in Ref. \cite{photon}.  In order to see this, let us consider the third and fourth terms of Eq. (\ref{L1}), 
\begin{equation}
{\mathcal{L}}_{int} =-\frac{1}{4}{\overline{\Psi}}\epsilon_{\mu\nu\rho\sigma}P\left[\beta^\mu,\;\beta^\nu\right]n^\rho\partial^\sigma\Psi+\frac{1}{4}{\overline{\Psi}} \epsilon_{\mu\nu\rho\sigma}P\left[\beta^\mu,\;\beta^\nu\right]n^\rho\overleftarrow{\partial}^\sigma\Psi.\label{3-4terms}
\end{equation}    If we consider the particular case where $n^\rho=\left(n^0,{\bf 0}\right)$, utilize $\left[\beta^i,\beta^j\right]=-i\epsilon^{ij}_{\ l}S^l$ (obtained from Eq. (\ref{spinoperator})), and use $PS_i=S_iP$, we observe that, for a complex field $\Psi$, the interaction term contains a spin-dependent structure,
\begin{equation}
{\mathcal{L}}_{int}=i\frac{n_0}{2}{\overline{\Psi}} \left({\bf  S}\cdot\nabla\right) P\Psi-i\frac{n_0}{2}{\overline{\Psi}}\left(\overleftarrow{\nabla} \cdot{\bf S}\right)P\Psi.\nonumber
\end{equation}

For the spin-one DKP field considered here, if we use the identity in Eq. (\ref{identity}) together with Eqs. (\ref{id-2}) and (\ref{id-4}), we find
\begin{equation}
{\mathcal{L}}_{int}=i\frac{n_0}{2}\overline{\Psi}\;{_\alpha R}{R^\alpha} \left({\bf  S}\cdot\nabla\right) \Psi-i\frac{n_0}{2}{\overline{\Psi}}\left(\overleftarrow{\nabla} \cdot{\bf S}\right){_\alpha R}{R^\alpha}\Psi,
\end{equation}
and from Eq. (\ref{spinoperator}) and the second identity in Eq. (\ref{prop-R}), we find
\bey
{\mathcal{L}}_{int}&=&i\frac{n_0}{2}{\overline{\Psi}} \left({ S}^l\partial_l\right){_i R}{R^i} \Psi-i\frac{n_0}{2}{\overline{\Psi}}{_i R}{R^i}\left(\overleftarrow{\partial}_l {S}^l\right)\Psi\nonumber\\
&=&-\frac{n_0}{4}{\overline{\Psi}} {_i R} \epsilon^l_{jk}\partial_l\left(g^{ki} {R^j }-g^{ji}R^k \right) \Psi+\frac{n_0}{4}{\overline{\Psi}}\epsilon^l_{jk}\overleftarrow{\partial}_l\left(g^{ki} {^j R}-g^{ji} {^k R}\right){R_i}\Psi\nonumber\\
&=&-\frac{n_0}{4}{\overline{\Psi}}{^k R} \left(\nabla \times{\bf R}\right)_k \Psi+\frac{n_0}{4}{\overline{\Psi}}{^k \left(\overleftarrow{\nabla} \times{\bf R}\right)}{R_k}\Psi,\nonumber
\eey
which, for a real field, reduces to
\begin{equation}
{\mathcal{L}}_{int}=-\frac{n_0}{2}{\overline{\Psi}}{\bf R} \cdot\left(\nabla \times{\bf R}\right) \Psi,\label{int=CS}
\end{equation}
which has the form of a Chern-Simons term, where ${\bf R}\Psi$ plays the role of the usual electromagnetic potential ${\bf A}$.

Hereafter, we shall utilize the following spin-one representation of the
beta matrices: 
\begin{equation}
\begin{array}{l}
\beta^0=e_{1,7}+e_{2,8}+e_{3,9}+e_{7,1}+e_{8,2}+e_{9,3}, \\ 
\beta^1=e_{1,10}+e_{5,9}-e_{6,8}+e_{8,6}-e_{9,5}-e_{10,1}, \\ 
\beta^2=e_{2,10}-e_{4,9}+e_{6,7}-e_{7,6}+e_{9,4}-e_{10,2}, \\ 
\beta^3=e_{3,10}+e_{4,8}-e_{5,7}+e_{7,5}-e_{8,4}-e_{10,3}. \\ 
\end{array}
\label{DKPs1}
\end{equation}
The (singular) gamma matrix for the massless term with a vector field is given by 
\begin{equation}
\gamma =3-\beta^\mu\beta_\mu=e_{1,1}+e_{2,2}+e_{3,3}+e_{4,4}+e_{5,5}+e_{6,6}.  \label{gamma}
\end{equation}
The shorthand notation $e_{ij}$ represents a $10\times10 $ matrix whose only
non-zero entry is $ij$, defined to be one, that is, $(e_{ij})_{mn} =
\delta_{im}\delta_{jn}$.

For the spin-one sector, we define the spinor $\Psi$ by 
\begin{eqnarray}
\Psi = \left( 
\begin{array}{c}
\Psi_1 \\ 
\vdots \\ 
\Psi_{10}
\end{array}
\right)= \left( 
\begin{array}{c}
{-i\mathbf{E}} \\ 
{-i\mathbf{B}} \\ 
{\mathbf{A}} \\ 
\phi%
\end{array}
\right)  \label{campo1}
\end{eqnarray}
where the electromagnetic fields are given by 
\begin{eqnarray}
{-i\mathbf{E}} = \left( 
\begin{array}{c}
\Psi_1 \\ 
\Psi_2 \\ 
\Psi_3 \\ 
\end{array}
\right),\;\;\;\; {-i\mathbf{B}} = \left( 
\begin{array}{c}
\Psi_4 \\ 
\Psi_5 \\ 
\Psi_6 \\ 
\end{array}
\right),\;\;\; {\mathbf{A}} = \left( 
\begin{array}{c}
\Psi_7 \\ 
\Psi_8 \\ 
\Psi_9 \\ 
\end{array}
\right)\;\;\;\phi=\Psi_{10}.  \label{campo2}
\end{eqnarray}
We can see that by identifying $\mathbf{E}$ as the electric field, $\mathbf{B%
}$ as the magnetic field, and $A_\mu=(\phi, \mathbf{A})$ as the usual
electromagnetic gauge field, then the free massless DKP equation above
reproduces the free Maxwell equations.

For many aspects, it is not necessary to recourse to a specific representation, by utilizing instead projection operators that select the wave-function for spin zero or spin one \cite{Umezawa, Lunardi2000}.  Hereafter, we shall proceed as in Ref. 
\cite{Lunardi2000} and construct operators of projection, 
\begin{equation}
R^\mu= (\beta^1)^2(\beta^2)^2(\beta^3)^2[\beta^\mu \beta^0 - g^{\mu 0}]
\label{op-r}
\end{equation}
and 
\begin{equation}
R^{\mu\nu}=R^\mu\beta^\nu,  \label{op-rr}
\end{equation}
that select the spin-one sector when applied to the DKP field $\Psi$. From
the operator properties in Eq. (\ref{prop-R}) of the Appendix \ref{AppB}, we see that
when we apply these operators on Eq. (\ref{dkp}), we find 
\begin{equation}
\partial_\mu\left(G^{\mu\nu}\Psi\right)=0,\qquad \partial_\mu\partial^\mu
\left(R^\nu\Psi\right)=0,  \label{proca}
\end{equation}
where 
\begin{equation}
G^{\mu\nu}\Psi=\partial^\mu (R^\nu\Psi)-\partial^\nu (R^\mu\Psi).
\label{U-eq}
\end{equation}
In other words, $R^\nu\Psi$ can be interpreted as a massless vector field
that satisfies the Klein-Gordon equation.

We can enforce a gauge symmetry which entails the interaction of the
spin-one DKP field $\Psi$ with an external gauge field $\mathcal{A}_\mu$.
Clearly, we can render the Lagrangian in Eq. (\ref{L}) invariant under some
gauge group G in the usual manner: given a gauge transformation $\Psi^{\prime }=S\Psi$, where $S$ belongs to G, we replace the partial
derivative $\partial_\mu$ with the covariant derivative $D_\mu=\partial_%
\mu-ig\mathcal{A}_\mu$, where $\mathcal{A}_\mu$ is a gauge field  that transforms as $\mathcal{A}_\mu^{\prime }=S\mathcal{A}_\mu S^{-1}-\frac
ig\left(\partial_\mu S\right)S^{-1}$. Note that $\mathcal{A}_\mu$ is different from the electromagnetic field $A_\mu$ described by the DKP field.

Since the first three terms of the Lagrangian in Eq. (\ref{L1}) correspond
to Eq. (\ref{L}), we are left to verify the gauge invariance of the last two
terms in Eq. (\ref{L1}). Let us modify $\frac{1}{4}{\overline{\Psi}}
\epsilon_{\mu\nu\rho\sigma}\left[\beta^\mu,\;\beta^\nu\right]%
n^\rho\partial^\sigma\Psi$ so that it becomes gauge-invariant under G (the
last term of Eq. (\ref{L1}) being treated the same manner). This is done
again by replacing the partial derivative $\partial_\mu$ by the covariant
derivative $D_\mu=\partial_\mu-ig\mathcal{A}_\mu$. We find that 
\begin{eqnarray}
D^{\prime \sigma}\Psi^{\prime }&=&\left(\partial^\sigma-igSA^\sigma
S^{-1}-(\partial^\sigma S)S^{-1}\right)S\Psi  \nonumber \\
&=&S\left(\partial^\sigma\Psi-igA^\sigma\Psi\right)  \nonumber \\
&=&SD^{\sigma}\Psi,  \nonumber
\end{eqnarray}
which shows the gauge invariance of $\frac{1}{4}{\overline{\Psi}}
\epsilon_{\mu\nu\rho\sigma}\left[\beta^\mu,\;\beta^\nu\right]n^\rho\partial^\sigma\Psi$, and hence of the Lagrangian in
 Eq. (\ref{L1}) when the gauge field is introduced with the covariant derivative.

The Lagrangian, Eq. (\ref{L1}), leads to the wave equation 
\begin{eqnarray}
\left( i\beta^\mu \partial_\mu -\frac{1}{2} \epsilon_{\mu\nu\rho\sigma}P \left[\beta^\mu,\;\beta^\nu\right]n^\rho\partial^%
\sigma- \gamma \right) \Psi=0  \label{dkp-int1}
\end{eqnarray}
and its adjoint equation, 
\begin{eqnarray}
{\overline{\Psi}}\left(i \beta ^{\mu}\overleftarrow{\partial}_\mu-\frac{1}{2}\epsilon_{\mu\nu\rho\sigma}P\left[\beta^\mu,\;\beta^\nu\right]n^\rho\overleftarrow{\partial}^\sigma+\gamma\right)=0.
\label{dkp-adjunta}
\end{eqnarray}

(Let us point out that if we wish to interpret Eq. (\ref{dkp-int1}) and its adjoint as describing a scalar field, then we must use projectors, $P_S$ and $P_\mu$, to select  the scalar sector of the DKP field, in analogy with Eqs. (\ref{op-r}) and (\ref{op-rr}) for the vector sector. However, when we apply these operators on Eq. (\ref{dkp-int1}), we can see that the Lorentz-breaking term disappears. As shown in the Appendix \ref{AppA}, for the scalar DKP field, $P_S[\beta^\mu,\beta^\nu]=0$ and $P_\mu(1-\gamma_S)=0$, so that the second term of Eq. (\ref{dkp-int1}) vanishes.  This corroborates the fact, mentioned earlier, that since the commutator $\left[\beta^\mu,\;\beta^\nu\right]$ of the new interaction term is related to the field's spin, then for spin-zero field, it will not contribute to the dynamics of the field.)

With the representations for $\beta^\mu$ and $\gamma$ given in Eqs. (\ref{DKPs1}) and (\ref{gamma}), we find that the DKP equation (\ref{dkp-int1}), modified by adding the background field $n^\mu$, takes the form 
\begin{eqnarray}
\mathbf{E}&=&-\nabla\phi-\partial_t \mathbf{A},  \nonumber \\
\mathbf{B}&=&\nabla\times\mathbf{A},  \nonumber \\
\nabla\times\mathbf{B}&=&\partial_t\mathbf{E}+n^0 \mathbf{B}+\mathbf{n}\times%
\mathbf{E},  \nonumber \\
\nabla \cdot\mathbf{E}&=&- \mathbf{n}\cdot\mathbf{B}.
\end{eqnarray}
One recognizes the Maxwell equations with the Coulomb and Amp\`ere laws
modified in terms of the background $n^\mu$ (see in Ref. (\cite{photon})).

Note that the energy-momentum tensor for the Lagrangian in Eq. (\ref{L1}) is 
\begin{eqnarray}
T^{\mu\nu}&=& \frac{i}{2} {\overline{\Psi}} \beta ^{\mu} \partial ^{\nu}
\Psi-\frac{i}{2} \left[ \partial ^{\nu} {\overline{\Psi}} \right] \beta
^{\mu} \Psi-\frac{1}{4}{\overline{\Psi}}\epsilon_{\lambda\alpha\rho\sigma}P\left[\beta^\lambda,\beta^\alpha\right]n^\rho
g^{\sigma\mu} \partial^\nu\Psi  \nonumber \\
& &+\frac{1}{4}\left[\partial^\nu{\overline{\Psi}}\right]\epsilon_{\lambda\alpha\rho\sigma}P\left[\beta^\lambda,\beta^\alpha\right]n^\rho g^{\sigma\mu}\Psi-g^{\mu\nu}{\mathcal{L}}.\label{Tmunum=0}
\end{eqnarray}
Note that the extensions, which contributes the last terms to Eq. (\ref{L1}), renders the energy-momentum tensor nonsymmetric: $T^{\mu\nu} \neq T^{\nu\mu}$ which again indicates the absence of Lorentz invariance. One can verify that $\partial_\mu T^{\mu\nu}=0$ by using the equations of motion in Eqs. (\ref{dkp-int1}) and (\ref{dkp-adjunta}), thus showing that  this tensor be conserved because the DKP theory is invariant under translations in the Minkowski space.

As shown in the Appendix \ref{AppC}, we can write Eq. (\ref{Tmunum=0}) as
\bey
T^{\mu\nu}=- {G}^{\alpha\mu} G^\nu{\;\alpha} +\frac{1}{4}g^{\mu \nu} {G}^{\sigma\alpha} G_{\sigma\alpha}+ \frac{n^\nu}{2}\epsilon^{\mu\alpha\rho\sigma}{G}_{\rho\sigma}R_\alpha\Psi,\label{Tmunum0}
\eey
so that, with the help of Eq. (\ref{campo2}), the components are
\bey
T^{00}&=&\frac12\left({\bf E}^2+{\bf B}^2\right)+\frac{n^0}{2} \left({\bf B}\cdot{\bf A}\right),\\
T^{0i}&=&\left({\bf E}\times{\bf B}\right)^i+\frac{n^i}{2} \left({\bf B}\cdot{\bf A}\right).
\eey
As expected, if we take $n^\mu=0$, these two expressions lead to the Maxwell field's energy density and the Poynting vector, respectively. The $n$-dependent terms are similar to the ones obtained in Ref. \cite{photon}.
	
We emphasize  that the minimal coupling used with this formalism opens a
window of  possibilities for the investigation of contributions that can arise in this context. We therefore investigated the possibilities of a
charged particle to bring information about the four-vector $n^{\rho }$ which violates
Lorentz symmetry via such minimal coupling ($n^{\rho }\partial ^{\sigma }\rightarrow n^{\rho }D^\sigma $), so that information can be obtained from experiments with interference phenomena and Berry phases (see in Ref. \cite{knutfase}). Another study would consist in investigating new ways to generate bound states (Landau levels \cite{knutlandau}) and new Berry phases with non-minimal couplings in this formalism.

%%%% 3 - Birefringence

\section{Birefringence  for the massless DKP field\label{birefringence}}

As mentioned earlier, vacuum birefringence may provide a sensitive
phenomenological signature of Lorentz-symmetry breaking and it was amply
studied (e.g. Carroll et all (1990) and Kostel\'ecky-Mewes (2008) in Refs. 
\cite{photon,photon1}, and \cite{coll-kost,RefBirefringence}). This is analogous to
the optical birefringence, associated to double-refraction, of anisotropic
materials that have an index of refraction that depends on the polarization
and the direction of propagation of light. Typically, the Maxwell equations
modified with a Lorentz-breaking term will lead to dispersion relations that
correspond to left-handed and right-handed modes. Hereafter we deduce a
similar effect with our Lorentz-violating extension of the DKP equation. 

If we multiply Eq. (\ref{dkp-int1}) by $R^\alpha$ followed by $R^{\alpha\delta}$, then we find 
\begin{eqnarray}
\left( i\partial_\mu R^{\alpha\mu} -\frac{1}{2} \epsilon_{\mu\nu\rho\sigma}R^{\alpha}P\left[\beta^\mu,\;\beta^\nu\right]n^\rho\partial^\sigma-R^{\alpha} \gamma \right) \Psi&=&0.  \label{dkp-int5}
\\
\left(iG^{\delta\alpha}-\frac{1}{2} \epsilon_{\mu\nu\rho\sigma}R^{\alpha\delta}P\left[\beta^\mu,\;\beta^\nu\right]n^\rho\partial^\sigma-R^{\alpha\delta} \gamma \right) \Psi&=&0.
\end{eqnarray}

If we use the identities (\ref{id-1})-(\ref{id-4}) presented in the
Appendix \ref{AppB}, together with Eq. (\ref{dkp-int5}), we obtain 
\begin{eqnarray}
\left( -\partial_\delta G^{\delta\alpha} -\frac{1}{2} \epsilon_{\mu\nu\rho%
\sigma}R^{\alpha}\left[\beta^\mu,\;\beta^\nu\right]n^\rho\partial^\sigma%
\right) \Psi=0.  \label{dkp-int8}
\end{eqnarray}

With the identity (\ref{id-15}) in the Appendix \ref{AppB}, the previous equation
becomes 
\begin{eqnarray}
\left( \partial_\delta G^{\delta\alpha} -\epsilon^{\alpha\nu\rho\sigma}
n_\nu\partial_{\rho}R_{\sigma}\right) \Psi=0.  \label{dkp-int12}
\end{eqnarray}

In order to obtain a dispersion relation between the frequency $\omega$ and
the wave-vector $\mathbf{k}$, we expand the field $\Psi$ in terms of plane
waves, $\Psi=\frac{1}{(2\pi)^4} \int \tilde{\Psi}(k) e^{ik_\mu x^\mu} d^4 k$,
with the four-vector 
\begin{equation}  \label{kfourvector}
k_\mu=\left(\omega, \mathbf{k}\right),
\end{equation}
and utilize Eq. (\ref{dkp-int12}), which leads to 
\begin{eqnarray}
\frac{1}{(2\pi)^4}\int \left[k_\mu k^\mu g^{\alpha \sigma}+k^\alpha
k^\sigma-i\epsilon^{\alpha\nu\rho\sigma} n_\nu k_{\rho}\right] R_\sigma\tilde{\Psi}(k)e^{ik_\mu x^\mu}d^4 k =0.  \label{campo}
\end{eqnarray}

If we utilize the Lorentz gauge condition, $\partial_\mu R^\mu \tilde{\Psi}(k)=0$, that is,  $k_\mu R^\mu\tilde{\Psi}(k)=0$, 
then we have 
\begin{eqnarray}
\left(k_\mu k^\mu g^{\alpha \sigma}-i\epsilon^{\alpha\sigma\nu\rho} n_\nu
k_{\rho}\right)R_\sigma\tilde{\Psi}(k)=0.  \label{birref-1}
\end{eqnarray}
When we mutiply this expression by $n_\alpha$, we obtain  $n^\sigma R_\sigma\tilde{\Psi}(k)=0$.
Then we multiply Eq. (\ref{birref-1}) by $k_\mu k^\mu
g_{\lambda\alpha}+i\epsilon_{\lambda\alpha\delta\gamma} n^\delta k^{\gamma}$%
, 
\begin{eqnarray}
\left[\left(k_\mu k^\mu\right)^2 g_{\lambda }^{\sigma}
-\epsilon_{\alpha\lambda\delta\gamma} \epsilon^{\alpha\sigma\nu\rho}
n^\delta k^{\gamma} n_\nu k_{\rho}\right]R_\sigma \tilde{\Psi}(k)=0.
\label{birref-4}
\end{eqnarray}
From the properties of the tensor $\epsilon_{\alpha\lambda\delta\gamma}$,
Eq. (\ref{birref-4}) leads to 
\begin{eqnarray}
\left[\left(k_\mu k^\mu\right)^2 +\left(k_\mu k^\mu\right)\left(n_\nu
n^\nu\right)-\left(n_\nu k^\nu\right)^2 \right]R_\lambda\Psi&=&0,
\label{birref-6}
\end{eqnarray}
from which it follows that 
\begin{eqnarray}
\left(k_\mu k^\mu\right)^2 +\left(k_\mu k^\mu\right)\left(n_\nu
n^\nu\right)-\left(n_\nu k^\nu\right)^2 =0.  \label{birref-7}
\end{eqnarray}

Let us choose $n_\mu=(n_0,0)$ and use Eq. (\ref{kfourvector}), so that Eq. (\ref{birref-7}) becomes 
\begin{eqnarray}\label{masslessdisprel}
\omega^4-2|\mathbf{k}|^2 \omega^2+|\mathbf{k}|^2\left(|\mathbf{k}|^2-n_0^2\right)=0,
\end{eqnarray}
which gives us the solution 
\begin{eqnarray}\label{mrdf}
\omega_{\lambda}=|\mathbf{k}|\big(1 + \lambda n_0/|\mathbf{k}| \big)^{1/2},
\end{eqnarray}
where $\lambda=\pm 1$ implies that the background $n^\mu$ splits the photons into two modes of propagation. This dispersion relation is similar in Refs. \cite{photon,photon1} and provides evidence for the violation of Lorentz invariance.

The dispersion relation, Eq. (\ref{mrdf}), leads to a modified group velocity,
\begin{eqnarray}\label{vg1}
v_{g} \equiv \frac{\partial \omega_{\lambda}}{\partial|\mathbf{k}|}= \frac{(1 + \lambda n_{0}/2|\mathbf{k}|)}{ \big(1 + \lambda n_0/|\mathbf{k}| \big)^{1/2} }.
\end{eqnarray}
This expression leads to rotations of the polarization of linearly polarized photons during their propagation (see, e.g. Ref. \cite{Mattingly}). The group velocity, $v_{g}( \lambda= +1)$, can exceed the speed of light, thereby introducing problems of causality (see also Ref. \cite{Reyes}). On the other hand, the phase velocity can be obtained with $v_{p}\equiv \omega_{\lambda}/{|\mathbf{k}|}$,
\begin{eqnarray}\label{vp1}
v_{p} = \big(1 + \lambda n_0/|\mathbf{k}| \big)^{1/2}.
\end{eqnarray}
Notice that the phase and group velocities are related through Rayleigh's formula: $v_{p}/v_{g} = 1 - \big( \omega_{\lambda}/v_{p}\big) \big(d v_{p}/d |\mathbf{k}| \big)$. Thus, from Eq. (\ref{vg1}) and Eq. (\ref{vp1}), we find
\begin{eqnarray}
\frac{v_{p} - v_{g}}{v_{g}} &=& \frac{\lambda n_{0}}{|\mathbf{k}| } \frac{1}{\big(1 + \lambda n_0/|\mathbf{k}| \big)^{1/2}}, \nonumber\\&=& \frac{\lambda n_{0}}{|\mathbf{k}|} + {\cal{O} }(n_{0}),
\end{eqnarray}
for large momenta, $|\mathbf{k}|$, such that  $n_{0}/|\mathbf{k}| << 1$. In the superluminal case, $\lambda=+1$, we have that $v_{p} > v_{g}$, a normal dispersion medium. In the subluminal case, $\lambda=-1$, this implies at $v_{g} > v_{p}$, an anomalous medium (from an influence of anisotropic effects). Therefore, we can conclude that a model truly isotropic, $(n_{\mu}\equiv (n_{0}, \bf{0}))$, must be attributed only to superluminal case. This is important for phenomenological analyses.

%%%%%%%%%   4 - The massive DKP spin-one field

\section{The massive DKP field with a background\label{massivedkp}}

As is well known, the existence of a massive gauge field implies that the associated
electromagnetic fields have short range, as we can see in a superconductor
environment (e.g. Meissner effect). Another example is the electroweak
theory where the vector bosons of the weak interaction acquire mass (which results in an
interaction confined within the atomic nucleus), whereas the long-range photon remains
massless. Hereafter, we observe that our model with the DKP equation
 reproduces previous results for the massive gauge fields with preferential spacetime directions which stem from the
violating background of the Carroll-Field-Jackiw term. Our dispersion relations is very similar to results found in the literature.

The Lagrangian associated to the spin-one sector of the DKP field with mass $m$ in a background $n^\nu$ is given by 
\bey
{\mathcal{L}} &=& \frac{i}{2} {\overline{\Psi}} \beta ^{\mu} \partial _{\mu}
\Psi - \frac{i}{2} \left( \partial _{\mu} {\overline{\Psi}} \right) \beta
^{\mu} \Psi -m{\overline{\Psi}} \Psi -\frac{1}{4m}{\overline{\Psi}} \epsilon_{\mu\nu\rho\sigma}P\left[\beta^\mu,\;\beta^\nu\right]n^\rho\partial^\sigma\Psi\nonumber\\
& &+\frac{1}{4m}{\overline{\Psi}} \epsilon_{\mu\nu\rho\sigma}P\left[\beta^\mu,\;\beta^\nu\right]n^\rho\overleftarrow{\partial}^\sigma\Psi.  \label{L1p}
\eey
 where $P=1-\gamma=\beta^\mu\beta_\mu-2$. 
 
 As we did with the massless DKP field (in Eq. (\ref{int=CS})), the interaction term
\begin{equation}
{\mathcal{L}}_{int} = -\frac{1}{4m}{\overline{\Psi}} \epsilon_{\mu\nu\rho\sigma}P\left[\beta^\mu,\;\beta^\nu\right]n^\rho\partial^\sigma\Psi+\frac{1}{4m}{\overline{\Psi}} \epsilon_{\mu\nu\rho\sigma}P\left[\beta^\mu,\;\beta^\nu\right]n^\rho\overleftarrow{\partial}^\sigma\Psi  \label{L2}
\end{equation}
also leads to a Chern-Simons type expression,
\bey
{\mathcal{L}}_{int}&=&-\frac{n_0}{4m}{\overline{\Psi}}{\bf R} \cdot\left(\nabla \times{\bf R}\right) \Psi+\frac{n_0}{4m}{\overline{\Psi}} \left(\overleftarrow{\nabla} \times{\bf R}\right)\cdot{\bf R}\Psi,
\eey
which for a real DKP field becomes
\bey
{\mathcal{L}}_{int}&=&-\frac{n_0}{2m}{\overline{\Psi}}{\bf R} \cdot\left(\nabla \times{\bf R}\right) \Psi. \label{int=CS-mass}
\eey
One should remember that the components of the DKP spinor $\Psi$ are multiplied by $m$, that is, ${\bf R}\Psi=m{\bf A}$.  This shows also that the analogous Eq. (\ref{int=CS}) for the massless DKP field cannot be obtained from Eq. (\ref{int=CS-mass}) simply by setting $m=0$.

From the Lagrangian in Eq. (\ref{L1p}), we obtain the equation of motion, 
\begin{eqnarray}
\left( i\beta^\mu \partial_\mu -\frac{1}{2m} \epsilon_{\mu\nu\rho\sigma}P\left[\beta^\mu,\;\beta^\nu\right]n^\rho\partial^\sigma- m \right) \Psi=0  \label{dkp-intp1}
\end{eqnarray}
and its adjoint 
\begin{eqnarray}
{\overline{\Psi}}\left(i \beta ^{\mu}\overleftarrow{\partial}_\mu-\frac{1}{2m}\epsilon_{\mu\nu\rho\sigma}P\left[\beta^\mu,\;\beta^\nu\right]n^\rho\overleftarrow{\partial}^\sigma+m\right)=0.
\label{dkp-adjuntap}
\end{eqnarray}
By performing a calculation similar to the massless case, we use the
representation in Eq. (\ref{DKPs1}) to obtain the Proca field equations
modified with the background  field $n^\mu$: 
\begin{eqnarray}
\mathbf{E}&=&-\nabla\phi-\partial_t \mathbf{A},  \nonumber \\
\mathbf{B}&=&\nabla\times\mathbf{A},  \nonumber \\
\nabla\times\mathbf{B}&=&\partial_t\mathbf{E}-m^2 \mathbf{A}+ n^0 \mathbf{B}+%
\mathbf{n}\times\mathbf{E}  \nonumber \\
\nabla \cdot\mathbf{E}&=&-m^2\phi- \mathbf{n }\cdot\mathbf{B}.
\end{eqnarray}

The calculations related to the energy-momentum tensor are also similar to
the massless case, 
\begin{eqnarray}\label{Tmunumass}
T^{\mu\nu}&=& \frac{i}{2} {\overline{\Psi}} \beta ^{\mu} \partial ^{\nu}
\Psi-\frac{i}{2} \left[ \partial ^{\nu} {\overline{\Psi}} \right] \beta
^{\mu} \Psi-\frac{1}{4m}{\overline{\Psi}}\epsilon_{\lambda\alpha\rho\sigma}P\left[\beta^\lambda,\beta^\alpha\right]n^\rho
g^{\sigma\mu} \partial^\nu\Psi  \nonumber \\
& &+\frac{1}{4m}\left[\partial^\nu{\overline{\Psi}}\right]\epsilon_{\lambda\alpha\rho\sigma}P\left[\beta^\lambda,\beta^\alpha\right]n^\rho g^{\sigma\mu}\Psi-g^{\mu\nu}{%
\mathcal{L}}.
\end{eqnarray}
With the help of the equations of motion (\ref{dkp-intp1}) and (\ref{dkp-adjuntap}), we find  $\partial_\mu T^{\mu\nu}=0$, as for the massless DKP field.  

As shown in Appendix \ref{AppC}, Eq. (\ref{Tmunumass}) can be expressed as Eq (\ref{Tmunumasscomplex}), and for $R^\mu\Psi$  a real field, the tensor $T^{\mu\nu}$ then becomes
\bey
T^{\mu\nu}=- {G}^{\alpha\mu} G^\nu{\;\alpha} +g^{\mu\nu}\left[\frac{1}{4}{G}_{\mu\alpha}  G^{\mu\alpha}+\frac{m^2}{2}\overline{\Psi}{^\alpha R}R_{\alpha}\Psi\right]+ \frac{n^\nu}{2}\epsilon^{\mu\alpha\rho\sigma}{G}_{\rho\sigma}R_\alpha\Psi.\label{Tmunumassreal}
\eey
Then we see that the components of this tensor are
\bey
T^{00}&=&\frac12\left({\bf E}^2+{\bf B}^2+m^2 A^\alpha A_\alpha\right)+\frac{n^0}{2} \left({\bf B}\cdot{\bf A}\right),\label{T00mass}\\
T^{0i}&=&\left({\bf E}\times{\bf B}\right)^i+\frac{n^i}{2} \left({\bf B}\cdot{\bf A}\right).\label{T0imass}
\eey
Clearly, with $m=0$ and $n=0$, we recover the Maxwell field's energy density and the Poynting vector, respectively.

The calculations related to the spatial anisotropy in this context are also
similar to the massless case and result in the expression 
\[
\left( k_{\mu }k^{\mu }-m^{2}\right) ^{2}+\left( k_{\mu }k^{\mu }\right)
\left( n_{\nu }n^{\nu }\right) -\left( n_{\nu }k^{\nu }\right) ^{2}=0. 
\]%
If we take $n_{\mu }=(n_{0},0)$ and $k_{\mu }=(\omega,\mathbf{k})$, we have 
\[
\omega^{4}-2\left( \vert\mathbf{k}\vert^{2}+m^{2}\right) w^{2}+\left( \vert\mathbf{k}\vert^{2}+m^{2}\right) ^{2}-\vert\mathbf{k}\vert^{2}n_{0}^{2}=0 , 
\]
which has solution 
\bey\label{MassDispersionRel}
\omega_{\lambda}=\left(\vert\mathbf{k}\vert^{2}+\lambda \left\vert \mathbf{k}\right\vert n_{0}+m^{2}\right)^{\frac12}, 
\eey
with $\lambda=\pm 1$, that expresses a spatial anisotropy effect for massive vector bosons. The term $\left\vert \mathbf{k}\right\vert n_{0}$ in Eq. (\ref{MassDispersionRel}) competes with the mass $m$ and this implies that the range of electromagnetic interaction depends on linear momentum; in particular, some specific values of $\mathbf{k}$ are such that the model would be effectively massless. Eq. (\ref{MassDispersionRel}) is analogous to Eq. (3.5) of  Ref. \cite{Andrianov} and its physical implications are similar to those discussed for the massless DKP field, after Eq. (\ref{masslessdisprel}).  The dispersion relation (\ref{MassDispersionRel})  carries information about the propagation modes of the theory.  Birefringence occurs when different polarization modes propagate at different (phase) velocities  than the wave propagates, thus determining a rotation in the polarization plane. In our case, in the massless limit, we find that there are two modes which propagate at different velocities, that is, we have the wave propagating with the two (right and left) modes of polarization with different velocities. Thereby we conclude that the background field promotes the birefringence of the model.

%%%  CONCLUSION

\section{Concluding remarks\label{conclusion}}

In this work we have examined the spontaneous violation of Lorentz symmetry
 for spin-one massless and massive bosonic fields with the DKP formalism. This is an
important question to verify what kind of contribution can emerge with this
formalism. In analogy with previous analyses of the CPT- and
Lorentz-symmetry breaking of the odd gauge sector of the Standard Model
Extension, we added similar symmetry-breaking terms in the
 DKP Lagrangians for spin-one massless and massive fields.
These symmetry-breaking terms, analogous to the Chern-Simons term
encountered in the literature, are defined in terms of the projection
operators of the DKP theory. Motivated by similar results in previous
studies, we verified that our model leads to dispersion and vacuum
birefringence effects. In particular, our results are compatible with previous calculations for the massless DKP field.
For both the massless and massive DKP theories, our
dispersion relations clearly show that the birefringence and anisotropy effect disappear
when the background field is equal to zero.

In this context, we can consider the appearance of Berry phases in the DKP equation and the study of anisotropies
generated by a background. Potential applications to the quantum phase include the use of a non-null field $n_i$ to study the Aharonov-Casher effect, by extending the CPT-even, dimension-five, non-minimal coupling between the Dirac and gauge fields of Ref. \cite{Casana2013} to a bosonic field described by the DKP equation. Also, along the lines of Ref. \cite{CarrollTam}, we could use the DKP formalism described in this paper to keep extra dimensions hidden by
 adding Lorentz-violating tensor fields, or aether, with expectations values aligned along the extra dimensions. In relation with dimensional reduction, let us cite the reduction from $3+1$ spacetime by means of only one space-like component $n_k$ of the background field with the DKP formalism in order to study the Chern-Simons interaction in $2+1$ spacetime.  Note also that the scenario described in this paper could be extended to use the DKP approach to describe scalar mesons, where the background field would be a tensor that violates Lorentz symmetry without  violating CPT.
 
One can also investigate the analogous
non-relativistic problems via the $5$-dimensional Galilean DKP theory, whose
results can be applied to phenomena in condensed matter physics. This is on
particular interest, given, for instance, the recent study on the broken
Galilean invariance at the quantum spin Hall edge \cite{Geissler} and in
spin-orbit coupled Bose-Einstein condensates \cite{Khamehchi}. 
Clearly, a natural continuation of the present work would be the introduction
 of the background field interaction term  within the Dirac Lagrangian.  

\section*{Acknowledgement}

E. Santos is grateful to the University of Alberta for its hospitality during his sabbatical leave.
We would like to express our thanks to the CNPq (Conselho Nacional de
Desenvolvimento Cient\'{\i}fico e Tecnol\'{o}gico, Brazil). M. de Montigny
acknowledges the Natural Sciences and Engineering Research Council (NSERC)
of Canada for partial financial support (grant number RGPIN-2016-04309).

\appendix

%%%%%%%%  Appendix A

\renewcommand{\theequation}{A.\arabic{equation}}

\setcounter{equation}{0}

\section{Appendix: DKP scalar projection operators \label{AppA} }

Following Refs. \cite{guertin} and \cite{Lunardi2000}, we define the operators $P_S$, $P_\mu$ and $\gamma_S$ as
\[
P_S=\frac13\left(\beta_\mu\beta^\mu-1\right),
\]
\[
P_\mu=P_S\beta_\mu,
\]
\[
\gamma_S=\frac 13\left(4-\beta_\mu\beta^\mu\right).
\]
They satisfy many properties, among which we need $P^\mu\beta^\nu=P_Sg^{\mu\nu}$ and $P_S^2=P_S$.  From the previous equations, we find $P_S(1-\gamma_S)=P_S$, so that when we apply $P_S$ to Eq. (\ref{dkp-int1}), the second term which contains $(1-\gamma_S)[\beta^\mu,\beta^\nu]$ becomes 
\[
P_S(1-\gamma_S)[\beta^\mu,\beta^\nu]=P_S[\beta^\mu,\beta^\nu]=0.
\]
We proceed in a similar way with $P^\mu$.  We can see that $P^\mu\gamma_S=P^\mu$, which implies $P^\mu(1-\gamma_S)$, so that the second term of Eq. (\ref{dkp-int1}) is also annihilated by $P^\mu$.

Therefore, since both projection operators $P_S$ and $P^\mu$ annihilate the second term of Eq. (\ref{dkp-int1}), then when we apply both operators to the equation of motion and combine the results, we see that $P_S\Psi$ satisfies the Klein-Gordon equation: $\partial_\mu\partial^\mu P_S\Psi=0$.

%%%%%%%%  Appendix B

\renewcommand{\theequation}{B.\arabic{equation}}

\setcounter{equation}{0}

\section{Appendix: DKP vector projection operators \label{AppB} }

From the algebra in Eq. (\ref{algebra2}), we have 
\begin{eqnarray}
\beta^\mu\beta^\nu\gamma=\gamma\beta^\mu\beta^\mu,  \label{id-1}
\end{eqnarray}
which implies 
\begin{eqnarray}
R^\mu \gamma&=&\gamma R^\mu=0, \quad R^\mu\left(1-\gamma\right)=R^\mu
\label{id-2} \\
R^{\mu\nu}\gamma&=& R^{\mu\nu},\quad R^{\mu\nu} (1-\gamma)=\gamma
R^{\mu\nu}=0,  \label{id-4}
\end{eqnarray}
We find also 
\begin{eqnarray}
R^{\alpha}\left[\beta^\mu\;\beta^\nu\right]=g^{\alpha\mu}R^\nu-g^{\alpha%
\nu}R^\mu  \label{id-15}
\end{eqnarray}

If we define 
\begin{equation}  \label{operator-r}
S^{\mu \nu} = \left[ \beta ^{\mu}, \beta ^{\nu} \right],  \label{operator-S}
\end{equation}
then we see that the operators defined in Eqs. (\ref{op-r}), (\ref{op-rr})
and (\ref{operator-S}) satisfy the following properties : 
\begin{equation}
\begin{array}{l}
R^{\mu\nu}=-R^{\nu\mu}, \\ 
R^\mu\beta^\nu\beta^\alpha=g^{\nu\alpha}R^\mu-g^{\mu\alpha}R^\nu, \\ 
R^{\mu\nu}\beta^\alpha=g^{\nu\alpha}R^\mu-g^{\mu\alpha}R^\nu, \\ 
R^\mu S^{\nu\alpha}=g^{\mu\nu}R^\alpha-g^{\mu\alpha}R^\nu, \\ 
R^{\mu\nu}
S^{\alpha\rho}=g^{\mu\rho}R^{\nu\alpha}-g^{\mu\alpha}R^{\nu\rho}+g^{\nu\alpha}R^{\mu\rho}-g^{\nu\rho}R^{\mu\alpha}. \label{prop-R}\end{array}
\end{equation}
Finally, let us note the following identity, which will be useful in Appendix \ref{AppC},
\bey
^\alpha R R_\alpha+\frac{1}{2} {^{\alpha\sigma} R} R_{\alpha\sigma}=1,\label{identity}
\eey
 where
\bey
^\alpha R=\eta \left(R^{\alpha}\right)^\dagger, \;\;\;\;\;\;\;{^{\alpha\sigma}R}=\eta \left(R^{\alpha\sigma}\right)^\dagger.
\eey

%%%%%%%%  Appendix C

\renewcommand{\theequation}{C.\arabic{equation}}

\setcounter{equation}{0}

\section{Appendix: Energy-momentum tensors\label{AppC}  }

In this appendix, we examine the energy-momentum tensors for the massive and massless DKP Lagrangians with a background field.  First, we consider the complex massive DKP field in more details, and then briefly highlight the analogous derivations for the massless field.

\paragraph{Complex massive DKP field.}

The massive DKP Lagrangian in Eq. (\ref{L1p}) leads to the wave equation
\bey
\partial_\mu G^{\mu\nu}+\frac{1}{2}\epsilon^{\mu\nu\rho\sigma}n_\mu G_{\rho\sigma}+m^2 R^\nu\Psi=0, \label{C1}
\eey
and its adjoint
\bey
\partial_\mu \overline{G}^{\mu\nu}+\frac{1}{2}\epsilon^{\mu\nu\rho\sigma}n_\mu \overline{G}_{\rho\sigma}+m^2 \overline{\Psi}{^\nu R}=0.  \label{C2}
\eey

We wish to express the components $T^{00}$ and $T^{0i}$ of the tensor $T^{\mu\nu}$ in terms of the components of $\Psi$. Consider the expression for $T^{\mu\nu}$: 
\begin{eqnarray}
T^{\mu\nu}&=& \frac{i}{2} {\overline{\Psi}} \beta ^{\mu} \partial ^{\nu}
\Psi-\frac{i}{2} \left[ \partial ^{\nu} {\overline{\Psi}} \right] \beta
^{\mu} \Psi-\frac{1}{4}{\overline{\Psi}}\epsilon_{\lambda\alpha\rho\sigma}P\left[\beta^\lambda,\beta^\alpha\right]n^\rho g^{\sigma\mu} \partial^\nu\Psi 
\nonumber \\
& &+\frac{1}{4}\left[\partial^\nu{\overline{\Psi}}\right]\epsilon_{\lambda\alpha\rho\sigma}P\left[\beta^\lambda,\beta^\alpha\right]n^\rho  g^{\sigma\mu}\Psi-g^{\mu\nu}{\mathcal{L}}.  \label{C3}
\end{eqnarray}
In order to find its components, we use its conservation ($\partial_\mu T^{\mu\nu}=0$) and the identity in Eq. (\ref{identity}). Note that 
\bey
m\overline{\Psi}\Psi&=&m\overline{\Psi}\left(^\alpha R R_\alpha+\frac{1}{2} {^{\alpha\sigma} R} R_{\alpha\sigma}\right)\Psi\nonumber\\
&=& m^2 A^\alpha A_\alpha+\frac{1}{2} \overline{G}^{\alpha\sigma} G_{\alpha\sigma}. \label{id1}
\eey

With Eq. (\ref{identity}), we see that the first term of the tensor $T^{\mu\nu}$ is
\bey
\frac{i}{2}\overline{\Psi}\beta^\mu\partial^\nu\Psi&=&\frac{i}{2}\overline{\Psi}\left(^\alpha R R_\alpha+\frac{1}{2} {^{\alpha\sigma} R} R_{\alpha\sigma}\right)\beta^\mu\partial^\nu\Psi\nonumber\\
&=&\frac{i}{2}\overline{\Psi}\left(_\alpha R R^{\alpha\mu}+\frac{1}{2} {_{\alpha\sigma} R} \left(g^{\sigma\mu}R^{\alpha}-g^{\alpha\mu}R^{\sigma}\right)\right)\partial^\nu\Psi\nonumber\\
&=&\frac{i}{2}\overline{\Psi}\left(_\alpha R R^{\alpha\mu}\right)\partial^\nu\Psi+\frac{i}{2} \overline{\Psi}\left({^{\alpha\mu} R}R_\alpha \right)\partial^\nu\Psi\nonumber\\
&=&-\frac{1}{2}\left(\overline{\Psi}_\alpha R\right) \partial^\nu G^{\mu\alpha}+\frac{1}{2} \overline{G}^{\mu\alpha} \left(\partial^\nu R_\alpha\Psi\right). \label{id3}
\eey
Likewise, the second term of the tensor $T^{\mu\nu}$ becomes:
\bey
-\frac{i}{2}\left(\partial^\nu\overline{\Psi}\right)\beta^\mu\Psi&=&-\frac{1}{2}\left(\partial^\nu \overline{G}^{\alpha\mu}\right) R_\alpha\Psi+\frac{1}{2}\left(\partial^\nu\overline{\Psi}{_\sigma R}\right)G^{\mu\sigma}, \label{id4}
\eey
so that by adding the expressions (\ref{id3}) and (\ref{id4}), we find
\bey
\frac{i}{2}\overline{\Psi}\beta^\mu\partial^\nu\Psi-\frac{i}{2}\left(\partial^\nu\overline{\Psi}\right)\beta^\mu\Psi&=&-\frac{1}{2}\left(\overline{\Psi}_\alpha R\right) \partial^\nu G^{\mu\alpha}+\frac{1}{2} \overline{G}^{\mu\alpha} \left(\partial^\nu R_\alpha\Psi\right)\nonumber\\
& &-\frac{1}{2}\left(\partial^\nu \overline{G}^{\alpha\mu}\right) R_\alpha\Psi+\frac{1}{2}\left(\partial^\nu\overline{\Psi}{_\sigma R}\right)G^{\mu\sigma}.\label{C-7}
\eey

Now it is crucial to add to the left-hand side of this expression (which is part of the tensor $T^{\mu\nu}$) the following zero-divergence term: 
\bey
K^{\mu\nu}=-\frac{1}{2}\left(\partial_\sigma\overline{\Psi} {^\nu R}\right){G}^{\mu\sigma}-\frac{1}{2} \overline{G}^{\mu\alpha} \left(\partial_\alpha R^\nu\Psi\right).\label{id5}
\eey 
Indeed, its divergence takes the form
\bey
\partial_\mu K^{\mu\nu}&=&-\frac{1}{2}\left(\partial_\mu\partial_\sigma\overline{\Psi} {^\nu R}\right){G}^{\mu\sigma}-\frac{1}{2}\left(\partial_\sigma\overline{\Psi} {^\nu R}\right)\partial_\mu {G}^{\mu\sigma}-\frac{1}{2} \left(\partial_\mu\overline{G}^{\mu\alpha}\right) \left(\partial_\alpha R^\nu\Psi\right)-\frac{1}{2} \overline{G}^{\mu\alpha} \left(\partial_\mu\partial_\alpha R^\nu\Psi\right)\nonumber\\
&=&-\frac{1}{2}\left(\partial_\sigma\overline{\Psi} {^\nu R}\right)\partial_\mu {G}^{\mu\sigma}-\frac{1}{2} \left(\partial_\mu\overline{G}^{\mu\alpha}\right) \left(\partial_\alpha R^\nu\Psi\right),
\eey
and from the equation of motion and the definitions $\tilde{{G}}^{\mu\nu}=\frac{1}{2}\epsilon^{\mu\nu\rho\sigma} {{G}}_{\rho\sigma}$, $\tilde{\overline{G}}^{\mu\nu}=\frac{1}{2}\epsilon^{\mu\nu\rho\sigma} {\overline{G}}_{\rho\sigma}$, we have
\bey
\partial_\mu K^{\mu\nu}&=&\frac{1}{2}\left(\partial_\sigma\overline{\Psi} {^\nu R}\right)n_\mu \tilde{G}^{\mu\sigma}+\frac{m^2}{2}\left(\partial_\sigma\overline{\Psi} {^\nu R}\right)R^{\sigma}\Psi+\frac{1}{2} n_\mu\tilde{\overline{G}}^{\mu\alpha} \left(\partial_\alpha R^\nu\Psi\right)+\frac{m^2}{2} {\overline{\Psi}}{^\alpha R} \left(\partial_\alpha R^\nu\Psi\right)\nonumber\\
&=&\epsilon^{\mu\sigma\alpha\rho}\left(\partial_\sigma\overline{\Psi}\right)n_\mu\left({^\nu R}  R_\rho- {_\rho R}  R^\nu\right)\left(\partial_\alpha\Psi\right)+\frac{m^2}{2}\left[\left(\partial_\sigma\overline{\Psi}\right) {^\nu R}R^{\sigma}\Psi+{\overline{\Psi}}{^\alpha R}  R^\nu\left(\partial_\alpha\Psi\right)\right].\nonumber\\
\eey
If we use a representation, we can see that ${^\nu R} R_\rho$ is different from zero only when $\nu=\rho$, which implies that the first term on the right-hand side becomes zero. Moreover, from this property, we can see that the term proportional to $m^2$ becomes zero when we perform the Fourier transforming for the fields $\Psi$ and $\overline{\Psi}$. These two conclusions imply  $\partial_\mu K^{\mu\nu}=0$. 

Then, we replace Eq. (\ref{C-7}) with
\bey
\frac{i}{2}\overline{\Psi}\beta^\mu\partial^\nu\Psi-\frac{i}{2}\left(\partial^\nu\overline{\Psi}\right)\beta^\mu\Psi&=&-\frac{1}{2}\left(\overline{\Psi}_\alpha R\right) \partial^\nu G^{\mu\alpha}+\frac{1}{2} \overline{G}^{\mu\alpha} G^\nu{\;\alpha}\nonumber\\
& &-\frac{1}{2}\left(\partial^\nu \overline{G}^{\alpha\mu}\right) R_\alpha\Psi+ \frac{1}{2}\overline{G}^{\nu}_{\;\alpha}G^{\mu\alpha} .
\eey
We note that the divergence of the first and the third terms on the right-hand side is
\bey
\partial_\mu\left[-\frac{1}{2}\left(\overline{\Psi}_\alpha R\right) \partial^\nu G^{\mu\alpha}-\frac{1}{2}\left(\partial^\nu \overline{G}^{\alpha\mu}\right) R_\alpha\Psi\right]&=&-\frac{1}{2}\left(\partial_\mu\overline{\Psi}_\alpha R\right) \partial^\nu G^{\mu\alpha}-\frac{1}{2}\overline{\Psi}_\alpha R \partial^\nu \partial_\mu G^{\mu\alpha}\nonumber\\
& &-\frac{1}{2}\left(\partial^\nu \partial_\mu\overline{G}^{\alpha\mu}\right) R_\alpha\Psi-\frac{1}{2}\left(\partial^\nu \overline{G}^{\alpha\mu}\right) \partial_\mu R_\alpha\Psi\nonumber\\
&=&\partial_\mu \left(\frac{g^{\mu\nu}}{4}\overline{G}_{\mu\alpha}  G^{\mu\alpha}+g^{\mu\nu}\frac{m^2}{2}\overline{\Psi}{^\alpha R}R_{\alpha}\Psi\right),
\eey
where we have used $\epsilon^{\alpha\mu\rho\sigma}{_{\sigma} R} R_\alpha=0$, since ${_{\sigma} R} R_\alpha$ is non-zero only when $\sigma=\alpha$. Then we can substitute
\bey
-\frac{1}{2}\left(\overline{\Psi}_\alpha R\right) \partial^\nu G^{\mu\alpha}-\frac{1}{2}\left(\partial^\nu \overline{G}^{\alpha\mu}\right) R_\alpha\Psi=g^{\mu\nu}\left[\frac{1}{4}\overline{G}_{\mu\alpha}  G^{\mu\alpha}+\frac{m^2}{2}\overline{\Psi}{^\alpha R}R_{\alpha}\Psi\right], \label{id6}
\eey
and
\bey
\frac{i}{2}\overline{\Psi}\beta^\mu\partial^\nu\Psi-\frac{i}{2}\left(\partial^\nu\overline{\Psi}\right)\beta^\mu\Psi&=&-\frac{1}{2} \overline{G}^{\alpha\mu} G^\nu{\;\alpha} - \frac{1}{2}\overline{G}^{\nu}_{\;\alpha}G^{\alpha\mu}\nonumber\\
& &+g^{\mu\nu}\left[\frac{1}{4}\overline{G}_{\mu\alpha}  G^{\mu\alpha}+\frac{m^2}{2}\overline{\Psi}{^\alpha R}R_{\alpha}\Psi\right].\nonumber\\
\label{id7}
\eey

Now we consider the third term of $T^{\mu\nu}$ in Eq. (\ref{Tmunumass}) and utilize the identity (\ref{identity}) to obtain
\bey
-\frac{1}{4}{\overline{\Psi}}\epsilon_{\lambda\alpha\rho\sigma}P\left[\beta^\lambda,\beta^\alpha\right]n^\rho g^{\sigma\mu} \partial^\nu\Psi&=&-\frac{1}{4}{\overline{\Psi}}\epsilon_{\lambda\alpha\rho\sigma}P\left[\beta^\lambda,\beta^\alpha\right]\left(^\delta R R_\delta+\frac{1}{2} {^{\delta\theta} R} R_{\delta\theta}\right)\nonumber\\
& & \times n^\rho g^{\sigma\mu} \partial^\nu\Psi \nonumber\\
&=&-\frac{1}{2}{\overline{\Psi}}\epsilon_{\lambda\alpha\rho\sigma}P\beta^\lambda\beta^\alpha\left(^\delta R R_\delta\right) n^\rho g^{\sigma\mu} \partial^\nu\Psi \nonumber\\
&=&\frac{1}{2}\epsilon^{\lambda\alpha\rho\mu}n_\rho{\overline{\Psi}} {_\alpha R}  R_\lambda  \left(\partial^\nu\Psi\right) \nonumber\\
&=&0. \label{id8}
\eey
We show in a similar manner that the fourth term of $T^{\mu\nu}$ on the right-hand side of Eq. (\ref{Tmunumass}) is also equal to zero.

Finally, let us add another divergenceless $n^\mu$-dependent term to the tensor $T^{\mu\nu}$, 
\bey
M^{\mu\nu}=\frac{n^\nu}{4}\epsilon^{\mu\alpha\rho\sigma}\overline{G}_{\rho\sigma}R_\alpha\Psi+\frac{n^\nu}{4}\epsilon^{\mu\alpha\rho\sigma}\overline{\Psi}{_\alpha R}G_{\rho\sigma}.\label{C-16}
\eey
Its divergence is given by
\bey
\partial_\mu M^{\mu\nu}&=&\frac{n^\nu}{4}\epsilon^{\mu\alpha\rho\sigma}\left(\partial_\mu G_{\rho\sigma}\right)R_\alpha\Psi+\frac{n^\nu}{4}\epsilon^{\mu\alpha\rho\sigma}G_{\rho\sigma}\left(\partial_\mu R_\alpha\Psi\right)\nonumber\\
& & +\frac{n^\nu}{4}\epsilon^{\mu\alpha\rho\sigma}\left(\partial_\mu \overline{\Psi}\right){_\alpha R}G_{\rho\sigma}+\frac{n^\nu}{4}\epsilon^{\mu\alpha\rho\sigma}\overline{\Psi} {_\alpha R} \left(\partial_\mu G_{\rho\sigma}\right)\nonumber\\
&=&\frac{n^\nu}{4}\epsilon^{\mu\alpha\rho\sigma}\left(\partial_{\rho}\overline{\Psi}\right){_\sigma R} R_\alpha\left(\partial_\mu\Psi\right)+\frac{n^\nu}{4}\epsilon^{\mu\alpha\rho\sigma}\left(\partial_\mu \overline{\Psi}\right){_\alpha R}R_\sigma \left(\partial_\rho \Psi\right)\nonumber\\
&=&0.
\eey

Thus, we have  for $T^{\mu\nu}$:
\bey
T^{\mu\nu}&=&-\frac{1}{2} \overline{G}^{\alpha\mu} G^\nu{\;\alpha} - \frac{1}{2}\overline{G}^{\nu}_{\;\alpha}G^{\alpha\mu}+g^{\mu\nu}\left[\frac{1}{4}\overline{G}_{\mu\alpha}  G^{\mu\alpha}+\frac{m^2}{2}\overline{\Psi}{^\alpha R}R_{\alpha}\Psi\right]\nonumber\\
& &+ \frac{n^\nu}{4}\epsilon^{\mu\alpha\rho\sigma}\overline{G}_{\rho\sigma}R_\alpha\Psi+\frac{n^\nu}{4}\epsilon^{\mu\alpha\rho\sigma}\overline{\Psi}{_\alpha R}G_{\rho\sigma}\label{Tmunumasscomplex}
\eey
with $\partial_\mu T^{\mu\nu}=0$. If we take $R^\mu\Psi$ to be a real field, then the tensor $T^{\mu\nu}$ becomes (\ref{Tmunumassreal}).

\paragraph{Complex massless DKP field.}  The Lagrangian for the massless DKP field, given in Eq. (\ref{L1}), differs from the massive Lagrangian in Eq. (\ref{L1p}) by the third term, with $m$ replaced by the singular matrix $\gamma$.  Note that, in the end, it turns out that the components $T^{00}$ and $T^{0i}$ for the massless field are simply obtained from Eqs. (\ref{T00mass}) and (\ref{T0imass}) with $m=0$.   However, hereafter we highlight the independent approach, analogous to the massive field,  starting with the Lagrangian in  Eq. (\ref{L1}) to highlight a few subtleties.  For instance, note that the wave equations for the field and its adjoint  can be obtained by setting $m=0$ in Eqs. (\ref{C1}) and (\ref{C2}), respectively.  However, the tensor $T^{\mu\nu}$ has the form in Eq. (\ref{C3}), the difference being hidden in the Lagrangian of the last term; this equation thus contains $\gamma$, both the massive and the massless fields. 

Let us point out that, for massless fields, Eq. (\ref{id1}) is replaced by
\bey
\overline{\Psi}\gamma\Psi&=&\overline{\Psi}\left(^\alpha R R_\alpha+\frac{1}{2} {^{\alpha\sigma} R} R_{\alpha\sigma}\right)\gamma\Psi\nonumber\\
&=&\frac{1}{2} \overline{\Psi}\left({^{\alpha\sigma} R} R_{\alpha\sigma}\right)\gamma\Psi\nonumber\\
&=&\frac{-i^2}{2} \overline{\Psi}\left(\overleftarrow{\partial^\sigma} {^\alpha R}-\overleftarrow{\partial^\alpha} {^\sigma R}\right)
\left(\overrightarrow{\partial_\sigma} { R_\alpha}-\overrightarrow{\partial_\alpha} { R^\sigma}\right)\Psi\nonumber\\
&=&\frac{1}{2} \overline{G}^{\alpha\sigma} G_{\alpha\sigma} \label{id1m0}
\eey
where we use again Eq. (\ref{identity}).  The expressions (\ref{C-7}), (\ref{id5}), (\ref{id8}) and (\ref{C-16}) are still valid for the massless field.

Thus, we obtain that for a complex massless DKP field, $T^{\mu\nu}$ is given by
\bey
T^{\mu\nu}=-\frac{1}{2} \overline{G}^{\alpha\mu} G^\nu{\;\alpha} - \frac{1}{2}\overline{G}^{\nu}_{\;\alpha}G^{\alpha\mu}+\frac{1}{4}g^{\mu \nu} \overline{G}^{\sigma\alpha} G_{\sigma\alpha}+ \frac{n^\nu}{4}\epsilon^{\mu\alpha\rho\sigma}\overline{G}_{\rho\sigma}R_\alpha\Psi+\frac{n^\nu}{4}\epsilon^{\mu\alpha\rho\sigma}\overline{\Psi}{_\alpha R}G_{\rho\sigma},\nonumber\\
\eey
and such that $\partial_\mu T^{\mu\nu}=0$. 

If we choose $R^\mu\Psi$ to be a real field, then the tensor $T^{\mu\nu}$ reduces to Eq. (\ref{Tmunum0}).

%%%%%%%%%%%%%%%%%%%%%%%%%%%%%%%%%

%%    REFERENCES

%%%%%%%%%%%%%%%%%%%%%%%%%%%%%%%%%

\end{document}